\documentstyle[aps,preprint,tighten,epsf]{revtex}

\newcommand{\bra}[1]{\left\langle{#1}\right|}
\newcommand{\ket}[1]{\left|{#1}\right\rangle}
\newcommand{\bracket}[3]{\bra{#1}#2\ket{#3}}
\begin{document}
\title{Instanton Effects in the Decay of Scalar Mesons}
\author{C.\,Ritter\footnote{E-mail: ritter@itkp.uni-bonn.de},
        B.\,C.\,Metsch\footnote{E-mail: metsch@itkp.uni-bonn.de},
        C.\,R.\,M\"unz\footnote{E-mail: muenz@itkp.uni-bonn.de},
        H.\,R.\,Petry
       }
\address{
         Institut f\"ur Theoretische Kernphysik,\\
         Universit\"at Bonn, Nu{\ss}allee 14-16, 53115 Bonn, Germany\\
         $\,$\\
         }
\date{\today}

\preprint{\vbox{Bonn TK-96-03 \hfill }}
\maketitle

\begin{abstract}
We show that instanton effects may play a crucial role in the decay of scalar
mesons into two pseudoscalars.  Particularly the branching ratios of two
meson decays of the $f_0(1500)$, which is considered as a glue-ball
candidate, are then compatible with an ordinary $q \bar{q}$-structure of this
resonance and a small positive SU(3) mixing angle, close to a result recently
calculated with the same instanton-induced force \cite{Kle95}.
\end{abstract} 
\narrowtext

\section{Introduction} \label{I}
Instanton effects seem to be outstandingly reflected in properties of scalar
and pseudoscalar mesons. An interaction induced by instantons
\cite{Hoo76,SVZ80} solves naturally the $U_A(1)$ problem for the $\eta - \eta'$
masses in the pseudoscalar sector \cite{Bla90,Mue94}.  In the framework of a
relativistic quark model, this interaction acting on scalar mesons has been
suggested to explain quantitatively the unusual mass spectrum of scalar mesons
roughly in terms of a low lying singlet and a higher lying octet of $q\bar{q}
$-states \cite{Kle95}.  Their peculiar strong decay pattern, however, has not
yet been consistently described.

\begin{figure}[htbp]
  \begin{center}
      \input{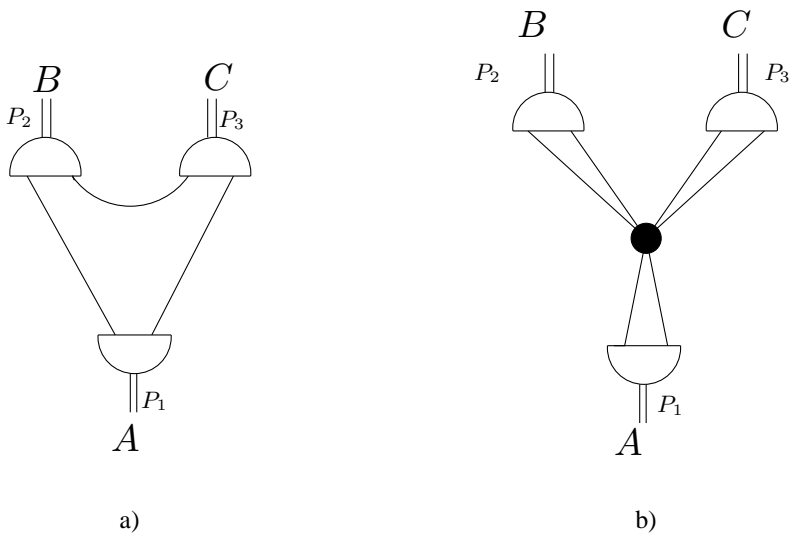}
  \end{center}
  \caption{{Feynman diagram of the strong decay of a meson in two
mesons \newline
a) in a conventional quark line diagram
b) by 't Hooft interaction}}
  \label{fig:Decay.pstex_t}
\end{figure}

In particular the $f_0(1500)$ \cite{cb} was argued to have properties
incompatible with a pure $q\bar{q} $ configuration and it was suggested to
possess a large glue admixture \cite{amslerclose}.  One of the major arguments
in favor of this interpretation is the decay phenomenology of the $f_0(1500)$:
It is found to decay into $\pi\pi \cite{gams},\; \eta\eta \cite{wa89},\;
\eta\eta^{\prime}$ \cite{toby} but not into $K\bar{K}$.  The $q\bar{q} $
hypothesis cannot fit {\it all} of these branching ratios at any $SU(3)_f$
scalar mixing angle, when decaying through a conventional decay mechanism (see
Fig.\ref{fig:Decay.pstex_t} in \cite{amslerclose}).  Furthermore the full width
$\Gamma(f'_0) =116\pm 17$~MeV is out of line with the scalar nonet: taking the
widths $\Gamma(a_0) = 270\pm 40$~MeV and $\Gamma(K_0^*) = 287\pm 23$~MeV as a
scale for the other members of the scalar nonet, a natural guess for the $f'_0$
is around 500 MeV. The $f_0(1500)$ thus does not naturally fit into the
quarkonium nonet.

In this paper we investigate the contribution of an instanton induced
six-quark-vertex to the strong decay of scalar mesons into pseudoscalars. We
derive $SU(3)_f$ branching ratios for the scalar octet, in particular for the
$f_0(1500)$, which allow for a region of the scalar mixing angle fulfilling all
of the three experimental branching ratios.  We also present results of a
numerical calculation in the framework of a relativistic quark model
\cite{Kle95}.

\section{The Instanton-Induced six-quark-vertex}
\label{II}
The flavor dependent effective quark interaction used here was computed by 't
Hooft and others from instanton effects \cite{Hoo76,SVZ80,Pet85}. 't Hooft
showed that an expansion of the (euclidian) action around the one instanton
solution of the gauge fields assuming dominance of the zero modes of the
fermion fields leads to an effective quark interaction not covered by
perturbative gluon exchange. For three flavors this is a six-point quark vertex
completely antisymmetric in flavor.  After normal ordering this results in a
contribution to the constituent quark masses, a two body interaction and a six
quark term that can be written as \cite{Mue94}:
\begin{eqnarray}
 & & \Delta{\cal L}{(3)}(y) =   \\ 
   & & \frac{27}{80}  g_{\mbox{eff}}^{(3)} 
      \Bigl\{
    :\! \overline{\Psi}(y) \, \overline{\Psi}(y) \,\overline{\Psi}(y) 
            \bigl[ 1 \! \cdot \! 1 \! \cdot \! 1
           \! +  \! \gamma_5 \! \cdot \!  
             \gamma_5 \! \cdot \! 1  
           \! + \!\gamma_5 \! \cdot \! 1
             \! \cdot \!  \gamma_5 
           \! + \! 1 \! \cdot \! \gamma_5
             \! \cdot \!  \gamma_5
      \bigr] 
       {\cal P}^F_1 (2{\cal P}^C_{10}+5{\cal P}^C_8)
       \Psi(y) \Psi(y) \Psi(y)\! :
      \Bigr\}    \nonumber 
\end{eqnarray}
where ${\cal P}^F_1$ is the projector onto a three-particle flavor singlet
state, ${\cal P}^C_{10}$ and ${\cal P}^C_8$ are projectors onto the color
decuplet and the color octet and $g_{\mbox{eff}}^{(3)}$ is an effective
coupling constant \cite{Mue94}. The dots imply that the first Dirac operator
acts on the first quark field and so forth.  Because of the special Dirac
structure the adequate formulation for this Lagrangian is the Weyl
representation for the Dirac spinors
\begin{equation}
  \Psi (x)=:{\xi (x) \choose \eta (x)}
\end{equation}
with diagonal $\gamma_5$. \\
The antisymmetric part of the operator acting on the three fermion fields
is given by the product of the antisymmetric flavor projector and a 
projector which is symmetric in spin and color. The only possible symmetric
spin-color operators are realized by the combinations:
\begin{eqnarray}
 {\cal P}_{1} & = & {\cal P}^{S}_{4} \otimes {\cal P}^{C}_{10} 
\nonumber\\
 {\cal P}_{2} & = & {\cal P}^{S}_{2} \otimes {\cal P}^{C}_{8} 
\end{eqnarray} 
with ${\cal P}^{S}_{4}$ the projector on the spin quadruplet and ${\cal
  P}^{Spin}_{2}$ the projector on the spin doublet.  The interaction Lagrangian
then can be written in terms of the Weyl spinors as:
\begin{equation}\label{thooft}
 \Delta{\cal L}{(3)} =    
   \frac{27}{20} \; g_{\mbox{eff}}^{(3)} 
      \,\Bigl\{
    \,:\, \eta^{\dag} \, \eta^{\dag} \,\eta^{\dag} 
           \,{\cal P}^F_1 \,(2{\cal P}_{1}+5{\cal P}_2)\,
       \xi \xi \xi\,:\,
      \Bigr\} + (\eta \longleftrightarrow \xi).
   \nonumber 
\end{equation}
We want to calculate the contribution of (\ref{thooft}) to the decay amplitude
of one meson into two mesons. In the present framework mesons are described as
bound $q \bar{q}$-states with a BS-Amplitude \cite{RMMP94}
\begin{equation}
 \chi^P_{J M_{Z}}(x,y) =
  \bracket{0}{T \Psi (x)\overline{\Psi}(y)}{P,JM_{Z}}
\end{equation}
It can be shown (see below) that the six-quark-interaction (\ref{thooft}) due
to its pointlike nature only applies to spin zero states i.e.  mesons with
nonvanishing Bethe--Salpeter amplitudes at the origin proportional to
$\gamma_5$ (for pseudoscalars) or $1$ (for scalars). Thus the only relevant
quantities for each meson are:
\begin{equation}\label{zwe}    
  \bracket{0}{T \xi (0)
    \eta^{\dag}  (0)}{P} \, \, \hbox{and} \,\,
   \bracket{0}{T \eta (0)
    \xi^{\dag}  (0)}{P} 
\end{equation}
Moreover the parity transformation relates these quantities:
\begin{equation} 
    \bracket{0}{T \xi (0)
    \eta^{\dag}  (0)}{P}   =  \pi \bracket{0}{T \eta (0)
    \xi^{\dag}  (0)}{P} 
\end{equation}
where $\pi$ is the parity of the meson.
The amplitude given in (\ref{zwe}) can be decomposed in a product of space, 
flavor, spin and color part:
\begin{equation}
 \bracket{0}{T \xi (0)
 \eta^{\dag}  (0)}{P} =  R(0) \otimes {\cal F} \otimes \Sigma \otimes {\cal C} 
\end{equation} 
The full 
contraction of the six-point Green's function 
\begin{equation}
{\cal G}^{(6)} = i\,\int d^4y\,
\left\langle\,0\,\left|\,T\,\Psi_{\alpha_2}\overline{\Psi}_{\beta_3}
\Psi_{\alpha_3}\overline{\Psi}_{\beta_2}\Psi_{\alpha_1}
\overline{\Psi}_{\beta_1}\,(\Delta{\cal L}{(3)} (y))\,
            \right|\,0\,\right\rangle
 \end{equation}

according to Wick's theorem
leads to the transition matrix element in lowest order for the decay process:
\begin{equation}\label{uberg}
\bracket{P_2 P_3}{\Delta {\cal L} (3)} {P_1} \,= \,
   \hbox{i} \, \frac{81}{5} \; g_{\mbox{eff}}^{(3)} 
 \; \; tr\bigl [6 {\cal P}^F_1 \,\bigotimes_{j=1}^3 \,(R(0) \otimes {\cal F})_j
 \bigr ]
   \; \;  tr\bigl [(2{\cal P}_{1}+5{\cal P}_2) \, \bigotimes_{j=1}^3 \,
  (\Sigma_j \otimes {\cal C}_j) \bigr] 
\end{equation}
where
\begin{equation} \label{not}
tr\bigl [{\cal O} \,\bigotimes_{j=1}^3 \, \chi^j \bigr ] =
{\cal O}^{\alpha_1 \alpha_2 \alpha_3 \beta_1 \beta_2 \beta_3 } 
 \; \chi^1_{\alpha_1 \beta_1} \; \chi^2_{\alpha_2 \beta_2} \; 
\chi^3_{\alpha_3 \beta_3}
\end{equation}
The second trace is non-vanishing only if all mesons have spin zero. It can
easily be calculated, since the mesons are color singlet states, and yields a
factor $\frac{160}{3 \sqrt{3}}$ (Color matrices are normalized by a factor
$\frac{1}{\sqrt{3}}$).  In the following section we will investigate the flavor
dependence of (\ref{uberg}).

\section{Decay Properties of Scalar Mesons in the SU(3) limit  }
\label{III}

The explicit flavor dependence in the notation of
equation (\ref{not}) is:
\begin{equation}
6 {\cal P}^F_1({\cal F}^1,{\cal F}^2,{\cal F}^3) := 
 {\cal F}^1_{ii'} \;{\cal F}^2_{jj'} \; {\cal F}^3_{kk'}
 \epsilon^{ijk} \; \epsilon^{i'j'k'}     
\end{equation}
with ${\cal F}^j_{kk'} \; j=1,2,3$ the flavor part of the meson amplitude j.
With the Cayley-Hamilton theorem this can be written as:
\begin{equation}
 6 {\cal P}^F_1({\cal F}^1,{\cal F}^2,{\cal F}^3)= tr({\cal F}^1 [{\cal
   F}^2,{\cal F}^3]_+) - 
(tr({\cal F}^1)tr({\cal F}^2 {\cal F}^2)+\hbox{cycl})+
tr({\cal F}^1)tr({\cal F}^2)tr({\cal F}^3)
\end{equation}
The first term is recognized as the flavor dependence of the conventional
quark line diagram (according to the OZI rule) of fig 1a).
Defining $\widehat{{\cal F}^l}$ as the traceless part of ${\cal F}^l$ we
obtain: 
\begin{equation}
 6 {\cal P}^F_1({\cal F}^1,{\cal F}^2,{\cal F}^3)= tr({\cal F}^1 [{\cal
   F}^2,{\cal F}^3]_+) - 
(tr({\cal F}^1)tr(\widehat{{\cal F}^2} \;\widehat{{\cal F}^2})+\hbox{cycl})
\end{equation}
The flavor dependence of the three-body interaction thus leads to a minimal
violation of the OZI rule: only if $tr({\cal F}^i)$ does not vanish, i.e. a
flavor singlet participates, there is an additional contribution to the
conventional decay mechanism given by the first term.

With a pseudoscalar mixing angle $\Theta_{PS}=-17,3^\circ$ for the $\eta
\eta'$-system \cite{Mix} we find the following partial widths (normalized to
$\Gamma(a_0 \longrightarrow K \bar{K})=1$) in the SU(3) limit:
\begin{eqnarray}\label{akst}
\Gamma(a_0 \longrightarrow \pi \eta ) & = & 0.4 \\
\Gamma(K^\ast \longrightarrow K \eta )  & = & 0.3 \nonumber \\ 
\Gamma(K^\ast \longrightarrow K \pi) & = & 1.5 \nonumber 
\end{eqnarray}

\begin{figure}[htbp]
  \begin{center}
      \input{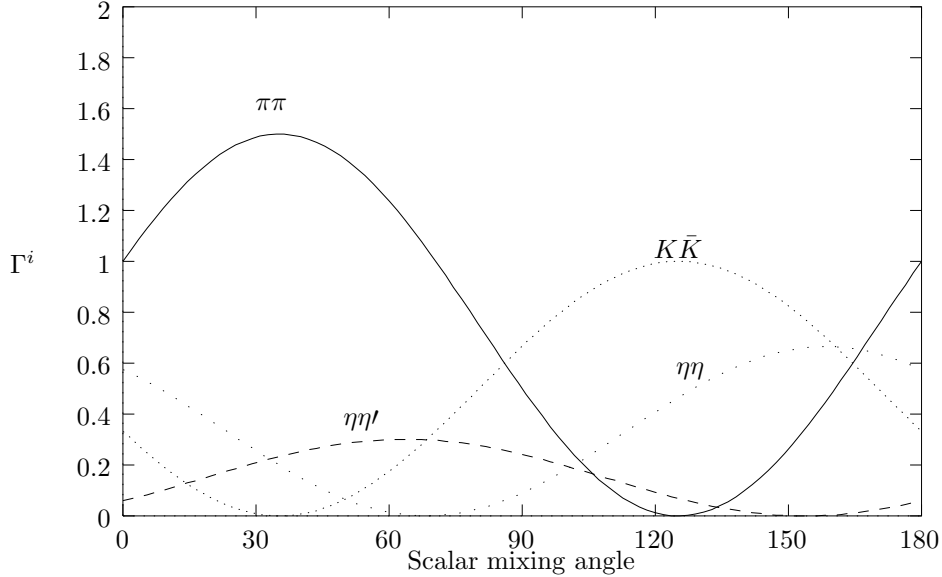}
  \end{center}
  \caption{{Relative partial width of an scalar 
isoscalar meson decaying into two pseudoscalars 
as a function of scalar mixing angle.}}
  \label{fig:fnulzerfall.pstex_t}
\end{figure}

Let us now consider the decay properties of the $f_0$, $f_0'$ system.
We parameterize the mixing of the singlet and octet $f_0$ as
\begin{eqnarray}
   \ket{f_0} & = & \sin(\Theta_{S}) \ket{f_{0,8}}
                  +\cos(\Theta_{S}) \ket{f_{0,1}} 
\\
   \ket{f_0'} & = & \cos(\Theta_{S}) \ket{f_{0,8}}
                  -\sin(\Theta_{S}) \ket{f_{0,1}} \nonumber
\end{eqnarray}
The partial widths of the $f_0'$ are plotted in 
Fig.\ref{fig:fnulzerfall.pstex_t} as a function of the scalar 
mixing angle $\theta_S$. With an angle of $\theta_S \approx 25^\circ$ we
find the following partial widths (normalized as in (\ref{akst})):
\begin{equation}
\pi \pi : \eta \eta :\eta \eta' :K \bar{K} = 1.45 
: \;\;\;\;\;0.32 \;\;\;\;\;: \;\;\;\;\;0.18 \;\;\;\;\;:  0.03
\end{equation}  
This is in fair agreement with the observed partial widths of the
$f_0(1500)$ \cite{amslerclose}:
\begin{equation}
\pi \pi : \eta \eta :\eta \eta' :K \bar{K} = 1.45:0.39\pm 0.15:0.28 \pm
0.12: <0.15
\end{equation}
if phase space and a form factor is divided out and the $\pi \pi$-width is
normalized.  It is interesting that in a relativistic quark model based on the
same instanton induced quark-antiquark force \cite{Kle95} we also find a small
positive mixing angle although the numerical value $ \theta \approx 6^\circ$ is
somewhat smaller. Though in this SU(3) model the total width of the $f_0'$ is
expected to be about as large as that of the $K^\ast$ this is still too large
to explain the remarkable small width of $\Gamma(f'_0) =116\pm 17$~MeV
\cite{fulw,fulww}.

\section{Instanton-induced scalar widths in a relativistic quark model}
\label{IV} 
The transition matrix elements for a scalar to two pseudoscalars have also been
calculated in the framework of a relativistic quark model based on the Salpeter
equation, for details see \cite{Kle95}.  There the mixing angle of the scalar
states is found to be $\theta \approx 6^{\circ}$ \cite{Kle95}.  As it stands,
the instanton-induced interaction is point-like.  We regularized the
interaction by replacing the delta function by a normalized Gaussian function,
which introduces a finite effective range \cite{Mue94}.  The calculation uses
all parameters as given in \cite{Kle95} and the calculated masses from this
model.  Our results for the partial widths are given in table \ref{tab}, where
we have adjusted the six-quark coupling strength $g_{\mbox{eff}}^{(3)}$ to the
process $K^\ast \longrightarrow K \pi$. Again, the small $f_0(1500)$ total
width cannot be accounted for. We infer that in future work we have to include
also the conventional decay mechanism of fig. 1a).  The interference between
both terms may well solve the remaining problems of the scalar decay modes.

For the $K^\ast$ the Particle Data Group \cite{PDG94} lists one decay mode
$K^\ast \longrightarrow K \pi$ within $93 \pm 10 \%$.  Our calculated result
($K^\ast \longrightarrow K \eta \approx 10\% $) is still compatible with this
number.  For the $f_0(980)$ there are two listed strong decay modes: $f_0(980)
\longrightarrow \pi \pi $ within $78.1 \pm 2.4 \%$ and $f_0(980)
\longrightarrow K \bar{K}$ within $21.9 \pm 2.4 \%$.  Our results, 60\% and
40\% respectively are still in fair agreement.

In the following we compare the invariant couplings for the $f_0(1500)$, 
i.e. the branching ratios divided by phase space factors:
\begin{equation}
\pi \pi :\eta \eta :\eta \eta' :K \bar{K} = 3 : \;\;\;\;\; \; 0.33 \;\;\;\;\;\;
:\;\;\;\;\;\;0.02\;\;\;\;\;\;:0.07
\end{equation}   
The experimentally seen invariant couplings \cite{cry95} are:
\begin{equation}
\pi \pi :\eta \eta :\eta \eta' :K \bar{K} = 
3 : 0.70\pm 0.27 :1.00\pm 0.46 : < 0.36
\end{equation} 
The clear discrepancy (for the $\eta \eta'$ channel) is mainly due to
the small calculated scalar mixing angle $\theta \approx 6^{\circ}$. 
The decay mechanism via the six-quark-vertex thus cannot explain
the experimentally seen branching ratios alone. The conventional contribution
has to be added coherently.
It is interesting to note  that the calculated invariant coupling
\begin{equation}
\Gamma(f_0(1500) \longrightarrow \pi \pi ):
\Gamma(f_0(1500) \longrightarrow \pi \pi(1300) ) = 1.2 : 1  
\end{equation}
indicates that the $\pi \pi'(1300) $-channel is relevant for the decay
of the $f_0(1500)$.

\section{Conclusion} \label{V}
In this paper an instanton induced six-point vertex has been worked out which
contributes to the strong decay of scalar mesons into two pseudoscalars. Using
a $SU(3)_f$ mixing angle $\theta \approx 25^{\circ}$ for the $f_0-f_0'$ system,
the experimental branching ratios of the $f_0(1500)$ can be explained in a
$SU(3)_f$ symmetric calculation in contrast to a mechanism via a conventional
quark line diagram. The small total width of the $f_0(1500)$, however, has not
yet been explained in the SU(3) calculation nor in a relativistic quark model.
It should be emphasized that none of the parameters was adjusted to the scalar
spectrum, and some improvement might be obtained by changing slightly the
strength of the instanton-induced force. In addition there is no logical
argument which rules out the conventional decay mechanism.  Actually the latter
still should account for all other meson decays, since the contribution
discussed here only works for (pseudo)scalars.  Therefore in the future the
interference between the conventional and the instanton-induced six-quark
vertex has to be worked out quantitatively.

\begin{table}
  \caption{Calculated decay widths in MeV with Bethe-Salpeter amplitudes
           from \protect{\cite{Kle95}} and
     instanton-induced six-quark-vertex}

  \label{tab}
  \centering
   \begin{tabular}{cccc}
  & $a_0(1450)\rightarrow\pi\eta$ & 101  &      \\
  & $a_0(1450)\rightarrow K K $ & 153   &      \\
  & $K_0^{\ast} (1430)\rightarrow K  \pi  $ & 264    &   \\
  &  $K_0^{\ast} (1430)\rightarrow K  \eta  $ & 28 &       \\
  & $ f_0(980)\rightarrow\pi\pi $ & 167 &     \\
  & $f_0(980)\rightarrow K K $ & 69 &      \\
  & $f'_0(1500)\rightarrow\pi\pi $ & 304  &      \\
  & $f'_0(1500)\rightarrow K  K $ &  5   &     \\
  & $f'_0(1500)\rightarrow\eta\eta$ & 21  &      \\
\end{tabular}
\end{table}

\end{document}